# Evaluation of a course mediatised with Xerte

Ghalia Merzougui[1], Roumaissa Dehkal[1], Maheiddine Djoudi[2].

[1] Dep.compute science, Faculty of science, University of Batna, Laboratory LASTIC
gmerzougui@live.fr, romaissau_314@hotmail.com
[2] XLIM-SIC, UMR CNRS 6172 and team IRMA laboratory
BD Marie and Pierre Curie - BP 3017986962 Futuroscope Chasseneuil Cedex France
mahieddine.djoudi@univ-poitiers.fr

**Abstract.** Interactive multimedia educational content has recently been of interest to attract attention on the learner and increase understanding by the latter. In parallel several open source authoring tools offer a quick and easy production of this type of content. As such, our contribution is to mediatize a course i.e. 'English' with the authoring system 'Xerte' which is intended both for simple users and developers in ActionScript. An experiment of course is conducted on a sample of a private school's students. At the end of this experience, we administered a questionnaire to evaluate the device, the results obtained, evidenced by the favorable reception of interactive multimedia integration in educational content.

## 1     Introduction

A multimedia course is an educational content that integrates and combines several media ie text, graphics, sound, image, animation and even video. Such content allows a new pedagogical approach with the use of more attractive methods where interactivity plays a big role. A course with such a characteristic allows to give a vision of knowledge that is different and complementary to that which is in a course called "classic". The learner, following a mediatized online course does not suffer but the runs, allowing it to stop or resume scrolling: The learner is more active in a mediatized online course, facilitating learning.

Richard Mayer, who developed the key concepts of "cognitive theory of multimedia learning" [1], tried to determine when learners can enjoy multimedia teaching materials to improve their assimilation and understanding. For example, students learn better from pictures and words put together with words put all alone; or when the audio (which represents the voice) and accompanying text are presented simultaneously, rather than a succession.

However, this training material is the result of a creative process and a series of well-defined actions. This process consists of five main parts: analysis, design, development, testing and distribution. Like any creative process, the most important work lies in the design. This step in turn contains three sub-steps including: structuring, Screenwriting and mediatisation.

[2] Bousbia tried to give a definition of the different stages. Structuring therefore for the cutting (hierarchical) of the knowledge in elementary units of very fine granularity. Screenwriting in turn is to give meaning to the hierarchical structure of content, organization and description of the transitions between the different concepts to be learned and planning for different sequences for each concept. The mediatization means: broadcast media content after structuring and screenwriting and definition of right media to integrate.

Several authoring systems have been developed which aims to facilitate a quick way the creation and diffusion of an interactive mediated courses. Our goal is therefore to mediatize a course namely "English: Introduction Level" using the Xerte authoring system. In the next section we present the services offered by this system that justifies our choice of its use. Then we try, in section three, to show our mediatization approach of the chosen course. Then, we present our analysis of the results (we find promising) obtained from an experiment proceeds realized within the learners of a private school that gives English language trainings.

## 2 The Authoring system Xerte

The Xerte is an open source tool with the permission of the University of Nottingham. [3] It allows the creation of courses with a variety of online interactive activities quickly and easily. With this tool, visually attractive and entertaining way for the user can be achieved in a relatively short time.

Using Xerte for the production of didactic sequences allows teachers to design their online courses themselves. Indeed, Xerte puts at their disposal sixty models (or template) predefined learning activities (synchronizing images with the video or audio, text exercises holes, drag & drop, scroll text, etc.). Xerte offers two ways to mediatize educational content, called by Xerte learning object or LO (Learning Object), namely Template method and tree method.

### 2.1 Template Method

In a single document or LO, you can create several pages (slides) where each slide is a template we chose. The figure below shows the seven categories namely Template : Texte, media, navigators, charts, interactivity, games and misc.

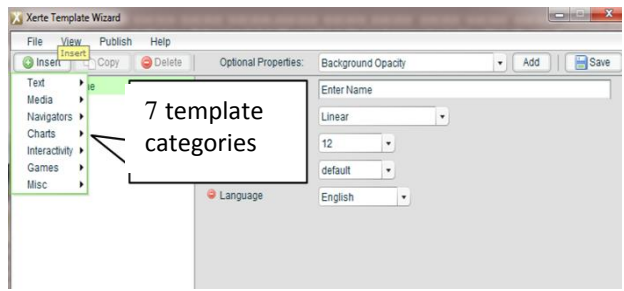

**Fig. 1.** Xerte interface in the Template mode

### 2.2 Tree Method

The document tree is composed of a set of pages on the left side of the interface see Fig.2. Each page can contain multiple pages and / or more media items from a range of eleven icons: text, picture, drawing, sound, video, script, Framework and interaction button, ... etc. The construction of the tree is done by dragging icons as needed by the user (add page, image, sound, ...).

In parallel with the Xerte authoring system, we used other tools for the creation and / or modification of media (audio, video, image, etc.); among them we mention: Mp3DirectCut, Movie maker, Wink, iSpring Free, etc.

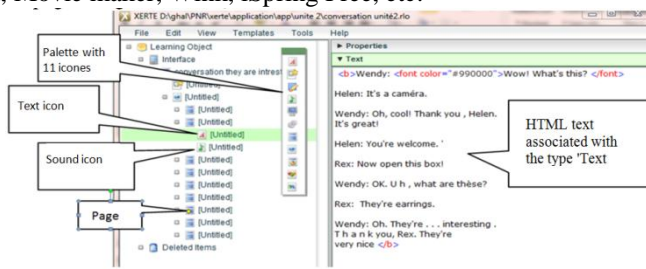

**Fig. 2.** Xerte interface in the tree mode.

## 3 English course mediatization

We tried to mediatize an English course taught in a private training school called "Elbadr for training" located in Batna. We will therefore follow the development process of an interactive multimedia application

### 3.1 Analysis

The course that was given to us by the school as a PDF document with a CD containing audio files, combines themes, functions and grammar. This course is structured in 16 units divided into several sections, each section has its own instant: Snapshot, Word Power, conversation, Pronunciation, Speaking, etc:

- **Snapshot** : generally introduces the topic of the unit with real information.
- **Word Power** : presents a new vocabulary.
- **Conversation** : is a dialogue that introduces a new grammar.
- **Pronunciation** : help to pronounce like a native speaker.
- **Speaking**: listen, you hear people talking in many contexts...

## 3.2 Course structuration

The analysis phase has shown us that the course is well structured into 16 units. Each unit covers several sections called learning activity. Now we start the screenwriting and the mediatization for each activity.

## 3.3 Screenwriting and mediatization

Learning Activity "**conversation"** (unit 1 section1) is a dialogue between two characters "Jennifer" and "Michael." the goal of this dialogue is to teach learner to present. In PDF support there is the text of the dialogue and an image of the two speakers (see Fig.3), the CD contains the audio conversation

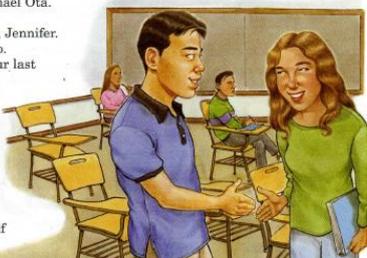

**Fig. 3.** Extract from the document English course [7]

We noticed that it is tedious for the learner to seek the audio file corresponding to the conversation that will listen and this reading the text on the book. Therefore, we thought to synchronize scrolling text with audio in the following scenario :

- The picture remains the same throughout the conversation.
- The synchronization between the display of colorful text and auditory pronunciation.
- Transit between the words is at the arrows, see Fig. 4.
- The click on the arrow next : the next word allows us to display with the red color and the preceding words become black and auditory pronunciation to play simultaneously.
- Previous arrow is back.

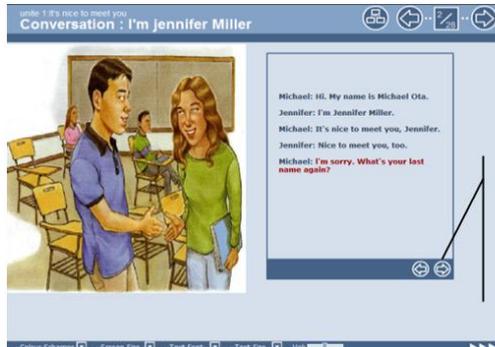

Arrows transit between words

**Fig. 4.** Activity « conversation » mediatised with Xerte

**Learning Activity : « pronunciation »** The objective of this activity is to tell the plural of the names and fill in the table to see Fig.5.

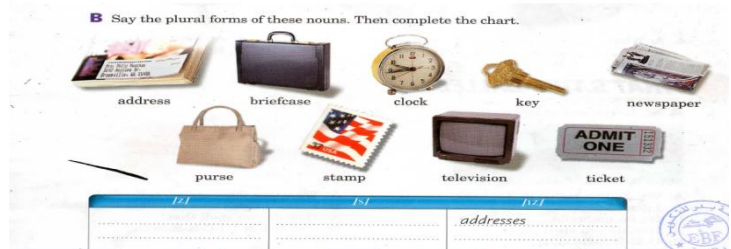

**Fig. 5.** Activity « pronunciation » Extract from [7]

We have chosen the template "class" of Xerte that allows us to determine the categories (in our case '/ Z', '/ S /', 'IZ') and the list of words to classify according to their pronunciation in the 3 categories . The result is shown in the figure below, where the list of words to classify is a set of labels (yellow rectangle) that can be moved by drag-drop with the mouse

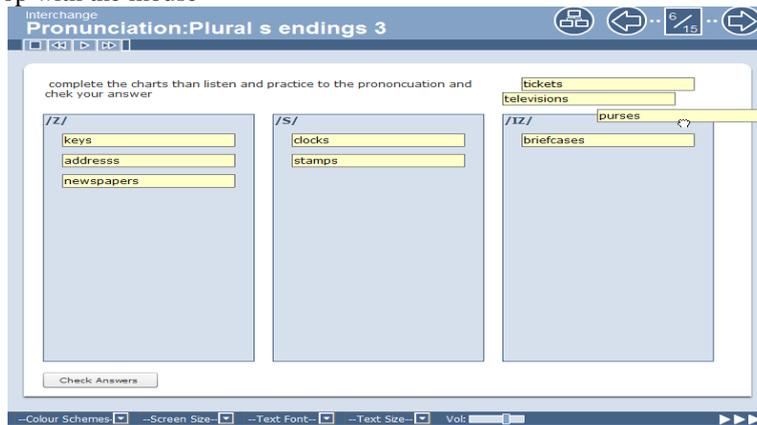

**Fig. 6.** Activity « pronunciation » mediatised with Xerte

**Learning Activity : « grammar focus »** This activity requires to complete the empty fields of a conversation with the appropriate text and practicing with a partner. The figure below shows the activity on the PDF support. We try to present how we are mediatised this activity.

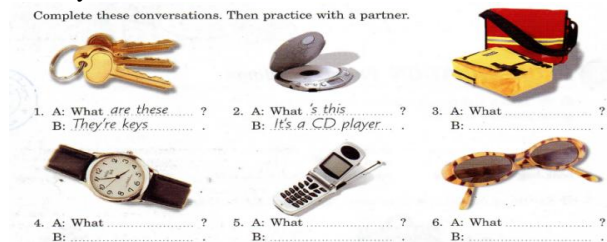

**Fig. 7.** « grammar focus » activity extract from PDF file[7]

For the media, we have included images of objects in Figure 7. To be interactive, we inserted under each image a component with type edit field. And to evaluate responses we've added buttons: one to display the score, another for display the correction and the last button for helps the learner to see some answers. We managed the actions of these buttons with ActionScript code associated with each of them (see Fig8).

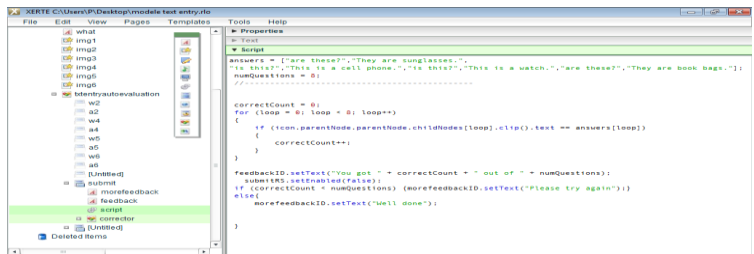

**Fig. 8.** activity « grammar focus » in Xerte tree mode.

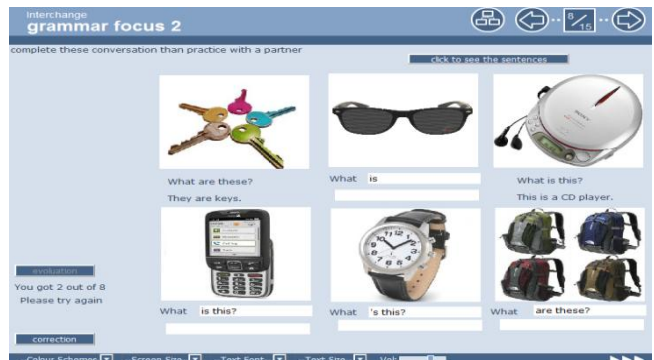

**Fig. 9.** « grammar focus» Activity médiatized with Xerte.

**Learning Activity: «Word Power »** The objective of this activity is to learn prepositions (in, behind, etc.). We thought to present this using interactivity. The learner has selected as a preposition (in, for example) in a list, the presentation shows him a picture that represents semantics (a key in a box for example). The author gives us a template system called "annotate diagram" that helps us to achieve this scenario. Fig 10 shows the final presentation of this activity.

.

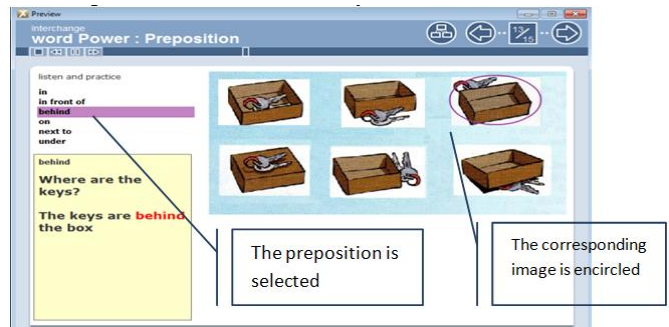
**Fig. 10.** Learnig activity « Word power : preposition» mediatized with Xerte.

**Learning Activity: «Listening spelling names »**

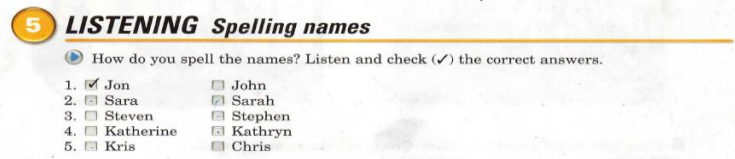
**Fig. 11.** Learning activity « Listening : Spelling names » extracted from PDF file [7]

This activity teaches the learner to distinguish well between letters and how to write their names. He must hear the audio then he ticks the right answer. With Xerte, we performed a presentation where the learner can listen to a part of the conversation (much times as he wants) and then selects one of the two proposals submitted by radio buttons. Then he can check his answer by clicking score button. He can go (back) to the following question (previous) by the arrows of transit.

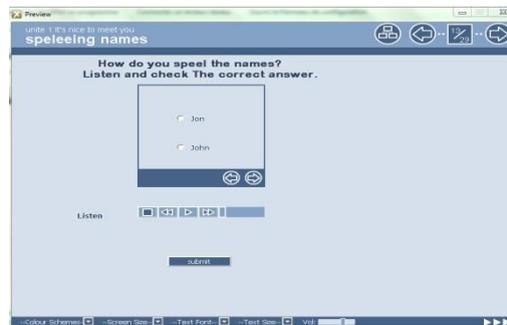
**Fig. 12.** L'activité « Listening Spelling names» médiatisé avec Xerte.

Among the 16 units, we have mediatized 3 units, 23 learning activities from Unit 1, 15 activities from Unit 2 and 6 activities from Unit 3.

## 4    Experimentation and evaluation

We conducted our experimentation with a sample consisting of 19 students from middle and secondary school who are enrolled in the private school "El Badr for training." This school provides training in the English language with an international method, and has an agreement with the Ambassador of the United States. In order to evaluate our interactive multimedia system, we developed an evaluation grid (see figure 13). To encourage participants to step back, we administered the anonymous evaluation questionnaire and face to face, just at the end of the experimentation. Learners disposed than 15 minutes to complete this survey. The following section presents the analysis and interpretation of the results obtained from this survey.

Objet : evaluation Quiz.

Objective: To evaluate an English course (level introduction) mediatised in flash

Dear students, dear students, please evaluate the course you will find on your PC using this questionnaire.
Give each proposal appreciation that matches your personal opinion. Please make a cross in the box that and leave blank boxes when the proposals are not appropriate

| | Good | Average | Bad |
|---|---|---|---|
| 1- The course is presented in a stimulating way | ☐ | ☐ | ☐ |
| 2- I appreciated the education received in the course. | ☐ | ☐ | ☐ |
| 3- The learning activities are useful to achieve the objectives of the course. | ☐ | ☐ | ☐ |
| 4- The synchronization of the text with his hearing pronunciation helps me to memorize the information. | ☐ | ☐ | ☐ |
| 5- Text Synchronization with his hearing pronunciation helps me to revise the lesson I Ratte. | ☐ | ☐ | ☐ |
| 6- The synchronization of the text with his hearing pronunciation helps me to train myself to say when I review myself. | ☐ | ☐ | ☐ |
| 7- The diversity of self-assessment activities break the routine of QCM and will not let me be ennui. | ☐ | ☐ | ☐ |
| 8- The reflection and deepening of matter are stimulated | ☐ | ☐ | ☐ |
| 9- I gain a lot of knowledge in the course. | ☐ | ☐ | ☐ |
| 10- The course encourages lean myself on content. | ☐ | ☐ | ☐ |
| 11- The course tries to convey enthusiasm. | ☐ | ☐ | ☐ |
| 12- The course is clearly structured. | ☐ | ☐ | ☐ |
| 13- The course is illustrated by examples | ☐ | ☐ | ☐ |
| 14- In comparison with the course (presential), your motivation level was high. | ☐ | ☐ | ☐ |
| 15- Formative evaluations lets me know my level of understanding | ☐ | ☐ | ☐ |
| 16- The media used to mediatise the course helps me to memorize the knowledge or concepts to grasp | ☐ | ☐ | ☐ |
| 17- The interactivity is enough to let me always attentive | ☐ | ☐ | ☐ |

comments
Your comments about the course and its relevance:
..................................................................................................................................
..................................................................................................................................

**Fig. 13. Evaluation grid**

## 4.1 Analysis and discussion of results

Statistics collected by questionnaires allow us to prepare an overall assessment of the experiment. The responses are used to compile the following charts:

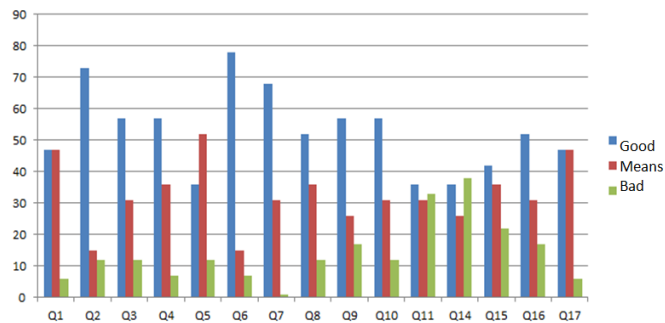
**Fig. 14.** Result obtained

The analysis of the results obtained from this experiment allowed us to summarize as follows:

Fromt the histogram above, we can conclude that according to the learners, the media education is well appreciated (Q2). On the other hand, students consider that the synchronization of text with it's ear pronunciation, helps a lot to train on pronunciation (Q6) and even on the information storage (Q4). Moreover, learners express their refusal of the substitution of the teacher in this type of content (Q14). We can deduce that more than half the learners (about 53%) are satisfied with the positive effect on their motivation, their attention (Q8, Q4). As it stimulates want them to look at the content and interact with it (Q9, Q10)5

## Conclusion

The results of this experimention shows the success of the implementation of a mediated interactive teaching. The participants in this experimentation were vast majority recognized the wealth of educational opportunities created by multimedia and interactivity. Here any time, some aspects that we consider important to highlight after our experiment:

We found that the inclusion of more activities prompting learning and interactive work, could encourage the learner to acquire a progressive way with some autonomy in his learning.

This work requires a large-scale evaluation through a questionnaire on an evaluation platform such as "SaWeb" for its valuation and determination of its impact on the learning process.

## References

1. R. Mayer, "Multimedia learning," Cambridge, MA, United States: Cambridge University Press (2001).
2. N. Bousbia, ''Contribution théorique et méthodologique à l'élaboration d'un environnement de FOAD'' Mémoire de magister, Institut National d'information INI, option système d'information, Alger, 2005.
3. J. Tenney, A. Beggan, ''Xerte online toolkits : content creation and distributed repository'' EDULEARN09 Proceedings, Pages 2280-2280, 2009, ISBN: 978-84-612-9801-3, ISSN: 2340-1117, 1st International Conference on Education and New Learning Technologies, le 6-8 July, 2009, Barcelona , Spain.
4. S. Ball, J. Tenney ''Xerte – A User-Friendly Tool for Creating Accessible Learning Objects'' Computers Helping People with Special Needs, Lecture Notes in Computer Science Volume 5105, 2008, pp 291-294
5. Clayton, John and Hall, Kevin "Using XERTE to create interactive learning materials". In: Shar-E-Fest 2013, 10-11 October, 2013, Hamilton, New Zealand.

6. E. El Bachir, E. Abdelwahed, M. El Adnani '' Projet d'innovation techno pédagogique dans l'enseignement secondaire au Maroc : retour d'expérience'' RADISMA, Numéro 6 (2010), 26 décembre 2010, http://www.radisma.info/document.php?id=1096. ISSN 1990-3219.
7. J. C. Richards '' Inetrchange" series 'Intro student's book' combridge university press, third edition, ISBN 0-521-60151-7, published 2005, www.cambridge.org.
8. M. Meloche, ''Evaluation des multimédias pédagogiques'', (2000) Consulté le 15 juin 2004 dans http://cqfd.teluq.uquebec.ca/D4_1_b.pdf
9. F. Gerard, (2003), '' L'évaluation de l'efficacité d'une formation'' *Gestion 2000*, Vol. 20, n°3, 13-33.